\documentclass[10pt,a4paper]{extarticle}
\pdfoutput=1

\usepackage[T1]{fontenc}
\usepackage[utf8]{inputenc}   
\usepackage[pdftex]{graphicx}                        
\usepackage[paper=a4paper, tmargin=1.75cm, bmargin=1.75cm, lmargin=1.25cm, rmargin=1.25cm, centering]{geometry}
\usepackage{mathpazo}
\usepackage{inconsolata}                             
\usepackage[hyphens,obeyspaces,spaces]{url}
\usepackage[svgnames]{xcolor}
\usepackage{multicol}
\usepackage{amsmath,amsfonts,amssymb,amsbsy}
\usepackage{setspace}
\usepackage{xspace}                                
\usepackage[detect-all]{siunitx}
\usepackage[font=small]{caption}
\usepackage{subcaption}
\usepackage{bm}
\usepackage{float}                                 

\allowdisplaybreaks                                   
\graphicspath{{figures/}}                           
\newcommand{\subrm}[1]{\ensuremath{_{\text{#1}}}}

\providecommand{\degree}{\ensuremath{^\circ}\xspace}    
\renewcommand{\degree}{\ensuremath{^\circ}\xspace}      
\newcommand{\silent}[1]{}
\newcommand{\figref}[1]{Fig.~\ref{#1}}

\newcommand{\vv}[1]{\bm{\mathrm{#1}}}                   

\newcommand{\um}{\,{\mu}\ensuremath{\mathrm{m}}\xspace} 

\makeatletter      
   \newcommand{\figcaption}{\captionsetup{type=figure} \def\@captype{figure}\caption}            
   \newcommand{\tabcaption}{\captionsetup{type=table} \def\@captype{table}\caption}            
   \renewcommand{\@biblabel}[1]{#1.}    
\makeatother

\def\citenum#1{\def\@cite##1##2{##1}\cite{#1}}

\begin{document}

\begin{center}
   {\huge\bf Raytracing and quantifying the inferior mirage} \\[18pt] 
   \textbf{Nathan Hagen}${}^{1,2*}$ and \textbf{Lucas Reyes Bentancor}${}^1$, \textbf{Kakeru Tanaka}${}^1$, \textbf{Norimasa Hanaya}${}^3$, \textbf{Momo Shoda}${}^3$, \textbf{Hiroki Maki}${}^3$, \textbf{Takuma Ando}${}^3$, \textbf{Keisaku Yoshizawa}${}^3$, \textbf{Kayo Suzuki}${}^3$, \textbf{Yume Yamada}${}^3$, \textbf{Asana Kumakura}${}^3$ \\[8pt]
   ${}^1$Utsunomiya University, Dept of Optical Engineering, 7-2-1 Yoto, Utsunomiya, Tochigi, Japan 321-8585 \\
   ${}^2$College of Optical Sciences, University of Arizona, 1630 E. University Boulevard, Tucson, Arizona 85721 \\
   ${}^3$Sakushin Gakuin High School, 1-1-41 Ichinosawa, Utsunomiya, Tochigi, Japan 320-0047
   ${}^{*}$email: nh@hagenlab.org
\end{center}

\vspace{1mm}
\setlength\columnsep{15pt}

\begin{abstract}
   Using custom raytracing code, we explore several features of the inferior mirage that are not widely discussed, such as the blurring in the mirage image, and what happens to the light in the direct image region that is replaced by the mirage image. The raytracing model also clarifies how to quantify the mirage, and we provide expressions to calculate the angular extent of the mirage under given thermal conditions and measurement geometry.
\end{abstract}

\begin{multicols}{2}

\section{Introduction}

The mirage is commonly used to illustrate light propagation through a gradient index medium. Its most familiar manifestation occurs when observing a long, straight road warmed by sunlight, producing a hot surface and cooler overlying air. Under these conditions, objects viewed at angles close to the horizon will appear to have a twin image reflected from the hot surface: the ``inferior mirage''.

While textbook illustrations do a good job at showing how geometric rays curve as they propagate through a gradient index medium, they are rarely accompanied by a quantitative analysis that would aid reader understanding~\cite{permissions}. The implied message is that a quantitative analysis is perhaps difficult, but this is unfortunate, since we show that it is neither difficult nor complex. In addition, conventional illustrations differ from models tied to realistic physical parameters, as we see in \figref{fig:textbook_mirages}, which shows an example illustration drawn from the current Wikipedia page on the mirage, and another derived from our ray tracing simulation. Differences include:
\begin{enumerate}
   \item Pedagogical illustrations of the mirage often show ray bending occurring at significant heights above the hot surface~\cite{Khular1977,Fabri1982}. For plausible physical situations, however, almost all of the ray bending occurs within millimeters of the surface~\cite{Croft1958,Minnaert1993,Puits2013,Ruiz2019,Wakid2019}. On the scale of the drawings, this tight bending would appear nearly the same as specular reflection from the road surface, so it is natural for authors to exaggerate the curving region for clarity. However, doing so without comment leaves the illustrations open to misinterpretation.
   \item Pedagogical drawings of the mirage generally show only a single ray emerging from an object point. However, for an image to form, there must be \emph{bundle} of rays from each object point such that each ray within the bundle converges to the same corresponding point on the image plane~\cite{Bush1974}.
\end{enumerate}
Moreover, because most discussions of the inferior mirage leave out quantitative models, it is easy for casual readers to receive the impression that the mirage only occurs in hot environments and strong sunlight. From physical principles, however, it is clear that the mirage must be present for \emph{any} thermal gradient, no matter how small. The real difference is that for small thermal gradients (e.g.\ $\Delta T < \SI{2}{\celsius}$), the angular extent of the mirage shrinks to the point that the visibility of the mirage becomes limited primarily by the height of observation, the flatness of the road, and having sufficient imaging magnification.

\begin{center}
   \includegraphics[width=0.9\linewidth]{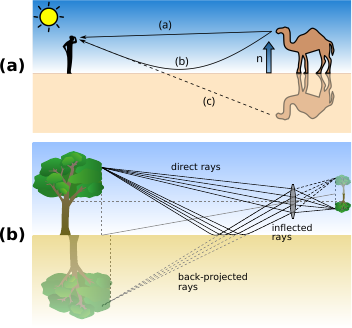}
   \figcaption{Example illustrations of the mirage, (a) from Wikipedia's page on the mirage~\cite{mirage_figure_from_wikipedia}, and (b) an illustration adapted from our simulation. Both figures use a squeezed horizontal axis for practicality.}
   \label{fig:textbook_mirages}
\end{center}

In the discussion below, we use a quantitative model of the inferior mirage and aim for improved physical accuracy through raytracing. Section~\ref{sec:raytrace} introduces three functionally equivalent ray tracing methods used for generating the geometric optical model: one based on solving the general form of the ray equation, another that uses the Snell law, and a third based on a closed-form expression. The source code for the calculations, and for generating all of the figures in the manuscript, is made available in the supplementary materials.
  
In Sec.~\ref{sec:model}, we build a quantitative model of the mirage thermal environment, and show how defocus and comatic aberrations arise to blur the mirage image. In Sec.~\ref{sec:inflection}, we continuing using the model to examine several little-explored features of the mirage, starting with an approximate formula for the inflection height for mirage rays, which quickly leads to an expression for calculating the angular range of the mirage (Sec.~\ref{sec:limits}). These provide a useful rule for estimating whether the mirage may be visible under given temperature conditions and viewing distances. 

Section~\ref{sec:measurements} shows experimentally captured photographs of an inferior mirage, which we analyze based on our earlier expressions. Closely looking at the mirage images brings up a curious question of what happens to the part of the scene replaced by the mirage image --- the light that is scattered directly from the hot pavement. We show that these scattered rays are deflected away from both the direct image and the mirage image as well, which is why we see a single reflected image rather than a reflection superposed on the underlying direct image.

\section{Ray tracing procedures}\label{sec:raytrace}

In order to simulate the mirage, we need to develop tools to accurately model light behavior in a gradient index medium. Towards this aim, we develop three equivalent methods: (1) solving the ray equation with finite difference equations, (2) using Snell's law in a discrete layer approximation, and (3) implementing the closed-form solution of Paola \textit{et al.}~\cite{Paola2022}. Each of these three approaches produces equivalent results.

The eikonal function $S$ of an optical system is proportional to the phase of the wavefunction, whose characteristic curves are light rays~\cite{Born2003short}. That is, light rays are everywhere orthogonal to the wavefronts of $S$, such that ray trajectories are the integral curves of the eikonal's gradient, $\vv{\nabla} S$. These properties allow one to express the path of rays in an arbitrary graded-index medium as solutions of the \emph{ray equation}~\cite{Sharma1982,Walther1995,Toro-Castillo2025}
\begin{equation}\label{eq:ray_equation}
   \vv{\nabla} n - \frac{d}{ds} \Big( n \frac{d \vv{r}}{ds} \Big) = 0 \, ,
\end{equation}
where $\vv{r}$ is the ray coordinate along the path $s$, and $d \vv{r}/ ds$ indicates the direction cosine vector of the ray. This can be manipulated into the form~\cite{Oliveira2019}
\begin{equation}\label{eq:ray_equation2}
   \frac{d^2 \vv{r}}{ds^2} = \frac{1}{n} \Big[ \vv{\nabla} n - \Big( \vv{\nabla} n \cdot \frac{d \vv{r}}{ds} \Big) \frac{d \vv{r}}{ds} \Big] \, . 
\end{equation}
The above term in the parentheses can be expanded as
\begin{equation}
   \Big( \vv{\nabla} n \cdot \frac{d \vv{r}}{ds} \Big) = \frac{\partial n}{\partial x} \frac{dx}{ds} + \frac{\partial n}{\partial y} \frac{dy}{ds} \, , 
\end{equation}
which when inserted into \eqref{eq:ray_equation2} leads to
\begin{equation}\label{eq:ray_eqn_computer_form}
   \left.
   \begin{aligned}
      \frac{d^2 x}{ds^2} &= \frac{1}{n}  \Big[ \frac{\partial n}{\partial x} - \Big( \frac{\partial n}{\partial x} \frac{dx}{ds} + \frac{\partial n}{\partial y} \frac{dy}{ds} \Big) \frac{dx}{ds} \Big] \, , \\
      \frac{d^2 y}{ds^2} &= \frac{1}{n}  \Big[ \frac{\partial n}{\partial y} - \Big( \frac{\partial n}{\partial x} \frac{dx}{ds} + \frac{\partial n}{\partial y} \frac{dy}{ds} \Big) \frac{dy}{ds} \Big] \, ,
   \end{aligned}
   \qquad \right\}
\end{equation}
where we assume $dz/ds = 0$ due to the symmetry of the mirage problem.

\begin{center}
   \includegraphics[width=0.95\linewidth]{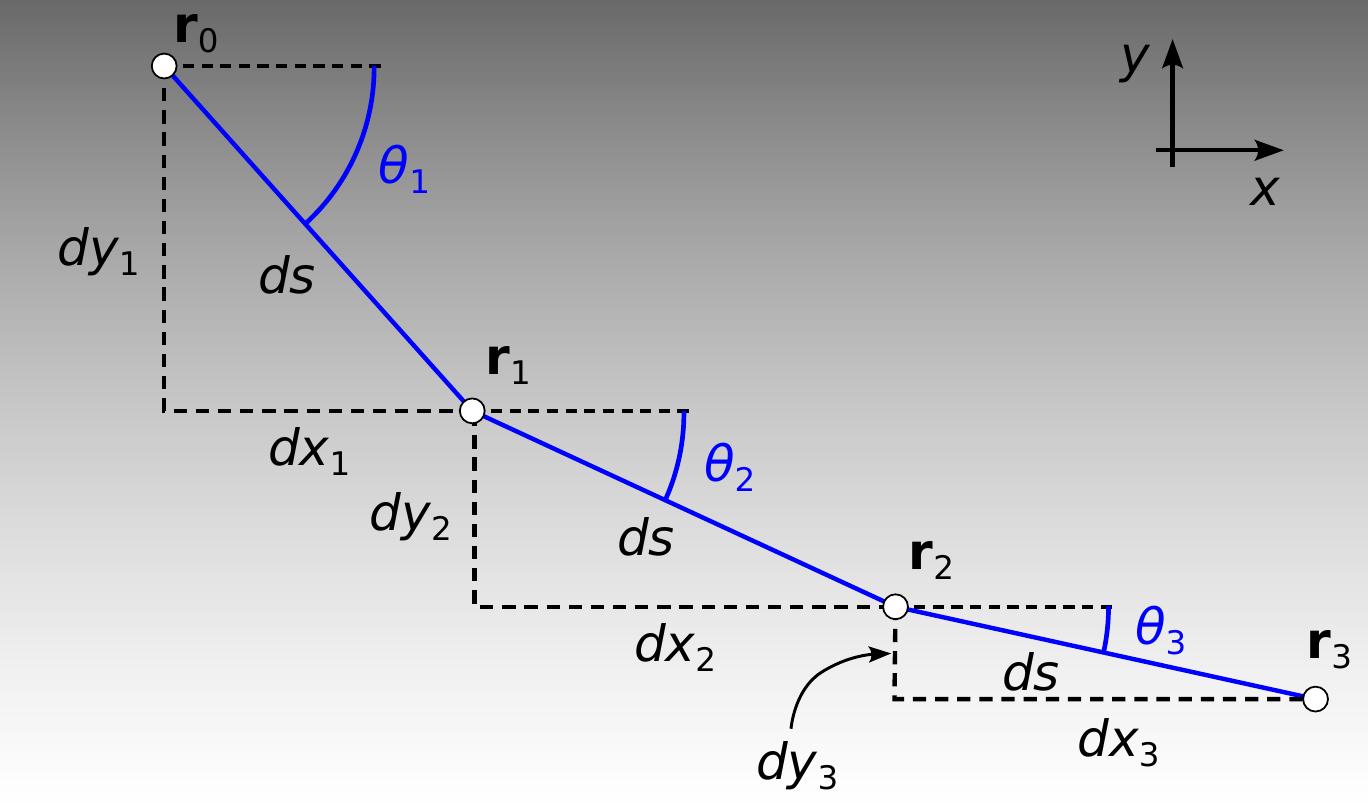}
   \figcaption{Ray-tracing procedure for solving the eikonal step-by-step through a gradient index medium.}
   \label{fig:eikonal_raytrace}
\end{center}
   
This last form is readily calculated by computer algorithms. At each given location $\vv{r}$ along a propagating ray, we estimate the local index gradient about the point, and use \eqref{eq:ray_eqn_computer_form} to calculate the curvature of the ray from the index gradient. This can be used as part of an iterative ray tracing procedure by selecting a starting point and direction for a ray, and also a stepsize. As illustrated in \figref{fig:eikonal_raytrace}, we take a small step along the ray's propagation direction and estimate the gradient vector at the new point. From the new gradient we calculate the ray curvature, which we use to update the ray direction vector. We then take another step and repeat the procedure. A useful advantage of this approach is that it allows any type of index gradient, as long as the stepsize is small enough that the refractive index change is small one point to the next.

Method \#2 for ray tracing the mirage uses the Snell law, while dividing the continuously varying refractive index into a series of discrete layers, as illustrated in \figref{fig:snell_raytrace}. For this we switch our angle reference from the horizontal (represented by $\theta$) to the surface normal (represented by $\phi$). At each interface between layers, we calculate the new ray angle from the Snell law
\begin{equation*}
   \phi_{i+1} = \arcsin \Big( \frac{n_i}{n_{i+1}} \, \sin \phi_i \Big) \, ,
\end{equation*}
after which we use this angle to propagate the ray to the next layer, find a new intersection point $x_{i+1},y_{i+1}$, and iterate. If we encounter a situation where the argument of the arcsine function is greater than 1, then we know that we have reached a total internal reflection. In this case, we flip the sign of the incoming angle, $\phi_{i+1} = -\phi_i$, and the sign of $dy$. Following this, we propagate to the next intersection and continue the same procedure.

\begin{center}
   \includegraphics[width=0.95\linewidth]{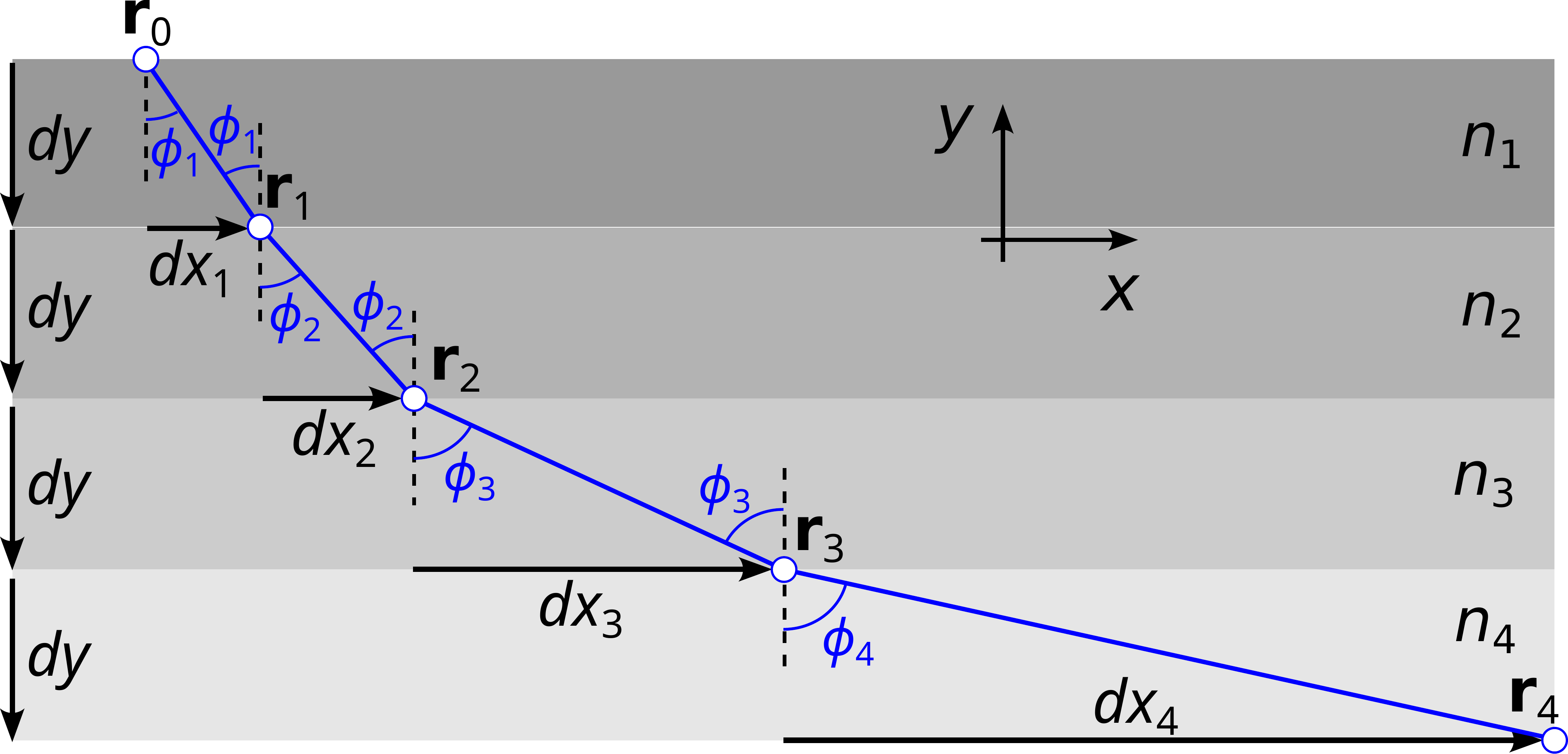}
   \figcaption{Step-by-step ray-tracing procedure using Snell's law at each discrete layer interface.}
   \label{fig:snell_raytrace}
\end{center}

It is easy to think of the Snell law procedure as reaching an impasse once it reaches a layer at which the propagation angle becomes strictly zero~\cite{Dyson1977}. This apparent difficulty has been treated in detail by Berry, who shows that it never in fact comes about~\cite{Berry2013}.

Method \#3 for calculating the ray trajectory is the closed-form equation provided by Paola \textit{et al.} provided in their remarkable set of three papers on the ray optics of mirages~\cite{Paola2022,Cruzado2023,Paola2024}. Introducing this approach requires describing the mirage environment model, and so we defer presenting it until Sec.~\ref{sec:inflection}.

\section{Mirage ray model}\label{sec:model}

For mild wind speeds, the average air temperature above an unconfined horizontal surface can be closely approximated as an exponential decay~\cite{Rotem1969},
\begin{equation}
   T (y) = T\subrm{cool} + (T\subrm{hot} - T\subrm{cool}) \exp (-y / \beta)
\end{equation}
where $y$ is the height above the surface, $\beta$ the thermal decay constant, $T\subrm{hot}$ the temperature of the surface, and $T\subrm{cool}$ the temperature of the air far above the surface. While many articles in the literature discuss the mirage and discuss quantitative models with ray trajectories, it is common to find example values given for $T\subrm{hot}$ and $T\subrm{cool}$, but remarkably few of them include the decay constant $\beta$. Of the two that we could find that do, Paola \textit{et al.} recommend using $\beta$ values in the range of 0.3--\SI{0.8}{cm} for measurements above pavement, while Horvath \textit{et al.} use $\beta = \SI{5}{cm}$ for measurements over water~\cite{Horvath2023,lapserate}.

Taking the Ciddor formula for the refractive index of air~\cite{Ciddor1996},
\begin{align*}
   m (\lambda) &= \frac{0.05792105}{\SI{238.0185}{\um^{-2}} - \lambda^{-2}} + \frac{0.00167917}{\SI{57.362}{\um^{-2}} - \lambda^{-2}} \, , \\
   n (\lambda,T) &= 1 + m (\lambda) \, \Big( \frac{\SI{288.15}{K}}{T} \Big) \, .
\end{align*}
In all of the simulations below, we assume a fixed wavelength of $\lambda = \SI{550}{nm}$. For plausible values of $\Delta T = T\subrm{hot} - T\subrm{cool}$, the change in refractive index with temperature is approximately linear, so that the refractive index distribution in air can also be approximated as an exponential,
\begin{equation}\label{eq:n_model}
  n (y) = n (T\subrm{cool}) - \big[ n (T\subrm{cool}) - n (T\subrm{hot}) \big] \, \exp (-y / \beta) \, .
\end{equation}
From this result, the centimeter-scale of the decay constant $\beta$ tells us that we can expect almost all of the ray bending to occur within a thin layer no more than a few centimeters thick above the warm surface.

Using this refractive index model, \figref{fig:sim2} shows a ray-based mirage simulation, where the road surface temperature is $T\subrm{hot} = \SI{49}{\celsius}$, the air temperature high above the surface is $T\subrm{cool} = \SI{43}{\celsius}$, and the thermal scale factor is set to $\beta = \SI{2.0}{cm}$, as it matches well with our experimental data. Rays emerge from an object point $y\subrm{obj} = \SI{100}{cm}$ above the road surface, and are viewed by a camera located at a distance of $x\subrm{lens} = \SI{600}{m}$. Because of the small angles involved, the vertical dimension is displayed at centimeter scale, while the horizontal is displayed in meters --- a pattern we repeat in all of our mirage figures below. The fan of 26 rays from the object extend from angles of $\theta \in \{ -0.2\degree, +0.02\degree \}$ with respect to the horizontal ($z$-axis), $\phi \in \{ 89.80\degree, 90.02\degree \}$ with respect to the downward-pointing $y$-axis. With a small value for $\beta$, and despite the magnified $y$-scale, the mirage rays still appear almost as if specularly reflected as from a mirror. However, in order to differentiate them from true surface specular reflections, we will use the word \textit{inflected} (rather than \textit{reflected}) to refer to rays that are smoothly bent away from the surface and form the mirage.

\begin{center}
   \includegraphics[width=0.95\linewidth]{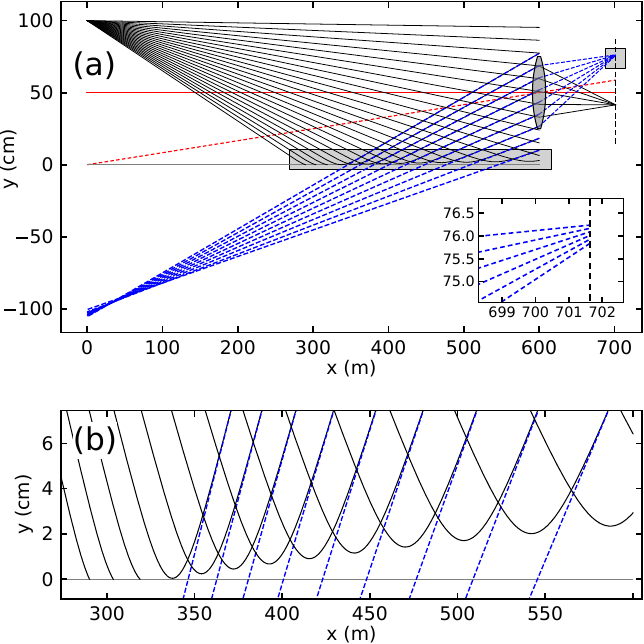}
   \figcaption{A mirage simulation, showing a ray fan of angles $\theta \in \{ -0.2\degree, +0.02\degree \}$ from an object point at height $y\subrm{obj} = \SI{100}{cm}$. Inflected rays that reach the lens plans are backprojected onto the object plane, and drawn in dashed blue lines. Those rays that enter the lens are focused onto the image plane, with the direct object rays shown in black, and the inflected object rays shown in dashed blue. The gray boxes indicate the inset region near focus and the magnified region shown in (b).}
   \label{fig:sim2}
\end{center}

In the figure, rays emerging from the object are drawn as black curves. Upon reaching the lens, direct rays are still drawn in black whereas inflected rays are drawn as dashed blue curves. Using each ray's angle and position at the lens, each inflected ray is also backprojected to the object plane. While rays that go directly from the object to the imaging lens without inflection appear to be straight, they are in fact very slightly curved due to the weak gradient refractive index at that height far above the surface. The direct and inflected rays that enter the lens are focused onto the image plane at the far right. The inset in \figref{fig:sim2}(a) shows a closeup view of the inflected ray bundle focus, where the spread of rays clearly indicate a modest amount of defocus.

\begin{figure*}
   \centering
   \includegraphics[width=0.85\linewidth]{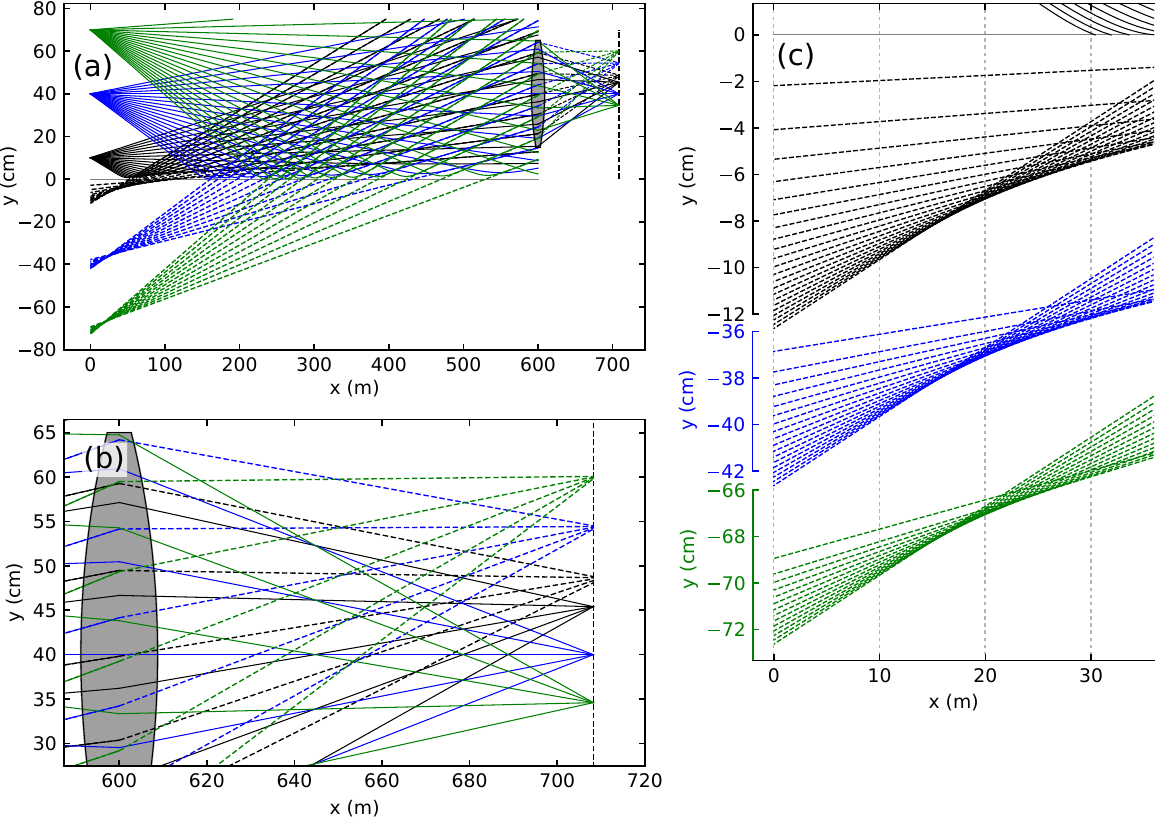}
   \caption{A mirage raytrace, (a) from object points at $y\subrm{obj}$ = 10, 40, and \SI{70}{cm}; (b) a closeup of the focal region; (c) a closeup of the mirage rays backprojected to the object plane.}
   \label{fig:sim3}
\end{figure*}

Rays leaving the object point with the largest (most negative) angles fail to curve sufficiently and hit the surface, where they are absorbed. These are the rays at the far left side of \figref{fig:sim2}(b). Of the rays that bend away from the hot surface, some of these enter the lens and focus onto the image plane. Rays that reach the lens plane outside of the lens aperture are truncated. Note that the size of the lens in the figure is greatly exaggerated for visual clarity. For the case shown here, the lens center height and diameter are $y\subrm{lens} = \SI{50}{cm}$ \& $D\subrm{lens} = \SI{50}{cm}$.

The focusing of the lens is modelled as an ideal paraxial lens, rather than as a real lens, such that the ray bending occurs at the midplane of the lens rather than twice at the front and real surfaces. This avoids any issues of possible image aberrations introduced by the lens itself. Thus, we can be confident that the defocus illustrated in the inset of \figref{fig:sim2}(a) is due to the inflection process and not to inaccuracies or aberrations in the lens model.

Figure~\ref{fig:sim2}(b) shows a closeup view of the regions where the rays bend away from the surface. The lowest point in each curve has been called the ``V-point'' due to its shape, and the locus of all V-points the ``inflection curve''. 

The red horizontal line in \figref{fig:sim2}(a) indicates the optical axis of the lens, which also indicates the location of the (infinitely distant) horizon, and its corresponding location on the image plane (if ignoring the Earth's curvature). The red dashed line indicates the approximate location where the inflected mirage image and the direct image touch.

Figure~\ref{fig:sim3} shows a mirage raytrace for a setup similar to that of \figref{fig:sim2} (i.e., $T\subrm{hot} = \SI{49}{\celsius}$, $T\subrm{cool} = \SI{43}{\celsius}$, $\beta = \SI{1.0}{cm}$, $x\subrm{lens} = \SI{600}{m}$) but where the object height varies as $y\subrm{obj}$ = 10, 40, and \SI{70}{cm}. Figure~\ref{fig:sim3}(b) shows a closeup of the focal region, where we can see that the defocus of the mirage rays actually increases as we approach the crossover point between the mirage and the direct image~\cite{Paola2024}. (The spread at the focus of black rays is larger than that for the blue and green rays.) This defocus difference may be large enough to become apparent when viewing under good viewing conditions.

Figure~\ref{fig:sim3}(c) shows a closeup view of the backprojected rays near the object plane. Similar to the effect at the image plane, smaller object heights produce larger blurring, such that the defocus and comatic aberrations in the backprojected rays become readily apparent. Berry~\cite{Berry2013} and Cruzado~\cite{Cruzado2023} provide detailed analyses of these aberrations.

\section{Ray inflection curve}\label{sec:inflection}

For modelling mirage rays, it is important to know the height at which a ray reaches its minimum height before deflecting upward --- the ray's ``V-point''. In order to locate this height $y_v$, we solve the Snell law for the refractive index at which the ray becomes horizontal, i.e.~\cite{Sastry1978}
\begin{equation}
   n_v = n\subrm{cool} \sin \phi_0 \, ,
\end{equation}
where $\phi_0$ is the initial ray angle at the object plane (calculated with respect to the $y$-axis). In order to get $y_v$, we invert the refractive index model \eqref{eq:n_model}:
\begin{equation}\label{eq:vpointy}
  y_v = - \beta \, \log \Big( \frac{n\subrm{cool} - n_v}{n\subrm{cool} - n\subrm{hot}} \Big) \, .
\end{equation}
Next, we can approximate the $x$-position of the V-point by ignoring ray bending, as if the mirage were generated by a flat mirror surface --- an approximation that is valid when $\beta$ is small with respect to the object height $y\subrm{obj}$. We therefore calculate the $x$-position from the starting ray height and angle, $y\subrm{obj}$ and $\phi_0$, and locate where the ray, if straight, would have intersected the surface:
\begin{equation}\label{eq:approx_vpointx}
   x_v \approx (y\subrm{obj} - y_v) \, \tan \phi \, .
\end{equation}
If needed, a more accurate formula is given by Eq.~(27) in Ref.~\citenum{Paola2022}:
\begin{equation}\label{eq:Paola_vpointx}
      \left.
   \begin{aligned}
      x_v &= 2 \beta \xi \, \ln \Big[ 1 + \Big( 1 - \exp [-(y\subrm{obj} - y_v) / \beta) \big] \Big)^{1/2} \Big] \\
         &\qquad + (y\subrm{obj} - y_v) \xi \, ,
   \end{aligned}
   \ \right\}
\end{equation} 
where
\begin{align*}
   \xi &= \frac{1 - \alpha \, \exp (-y_v / \beta)}{\sqrt{2 \alpha} \, \exp [-y_v / (2\beta)]} \, , \\
   \alpha &= 1 - n\subrm{hot} / n\subrm{cool} \, .
\end{align*}
Figure~\ref{fig:inflection} shows the resulting inflection curves and V-points calculated using both the accurate and approximate $x$-values, for $\beta$ values from 0.15 to \SI{0.9}{cm}. The inflection points tend to occur at heights $y \lesssim 3 \beta$.

\begin{center}
   \includegraphics[width=0.95\linewidth]{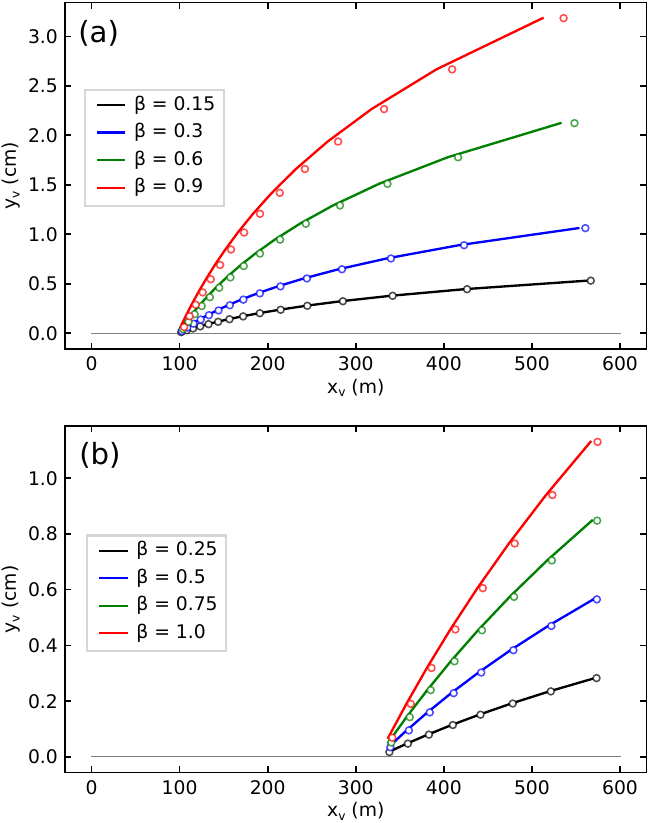}
   \figcaption{The inflection curves for the mirage rays, showing the $(x_v,y_v)$ coordinates for (a) $y\subrm{obj} = \SI{30}{cm}$ and (b) $y\subrm{obj} = \SI{100}{cm}$. The circles indicate the accurate V-points obtained from with Eqs~\ref{eq:vpointy}\,\&\,\ref{eq:Paola_vpointx}, while the curves indicate the approximate V-points obtained with Eqs~\ref{eq:vpointy}\,\&\,\ref{eq:approx_vpointx}.}
   \label{fig:inflection}
\end{center}

\section{The angular limits of the mirage}\label{sec:limits}

For any given environmental situation, it would be useful to know the angular extent of the mirage, in order to estimate its visibility. This is equivalent to asking the angular extent of the inflected ray bundle that reaches the lens, as drawn in \figref{fig:sim2}. While the lens in \figref{fig:sim2} is drawn at an exaggerated size, practical camera lenses ($D\subrm{lens} < \SI{75}{mm}$) and human eye pupils ($D\subrm{lens} < \SI{8}{mm}$) will generally be small relative to $y\subrm{lens}$. This allows us to use the chief ray passing through the center of the lens to define the limit of visibility. If the inflected ray reaches the lens significantly higher than $y\subrm{lens}$, then it will not be captured by the lens to reach the image, as indicated by the gray ray drawn in \figref{fig:min_max}. Thus, this ray defines the upper range of object heights that are visible.

Next, we assume that the lens is much higher than the primary region of the thermal gradient, i.e.\ $y\subrm{lens} \gg \beta$, and that the ray bending region is small enough to approximate the mirage ray as a specular reflection from the surface, i.e.\ $y\subrm{obj} \gg \beta$. Figure~\ref{fig:min_max} shows the geometrical layout for the calculations. The maximum ray angle possible for the mirage occurs where the object height is zero --- indicated by the red ray as $\theta_{+}$. From \figref{fig:min_max}, we see that this angle is given by
\begin{equation}\label{eq:theta_plus}
   \theta_{+} \approx \arctan(y\subrm{lens} / x\subrm{lens}) \, ,
\end{equation}
so it is governed by the camera setup alone, and is not a direct property of the mirage. This is also the angle where the real and mirage images meet (the midpoint at the image plane between the black ray focus and blue ray focus).

\begin{center}
   \includegraphics[width=0.7\linewidth]{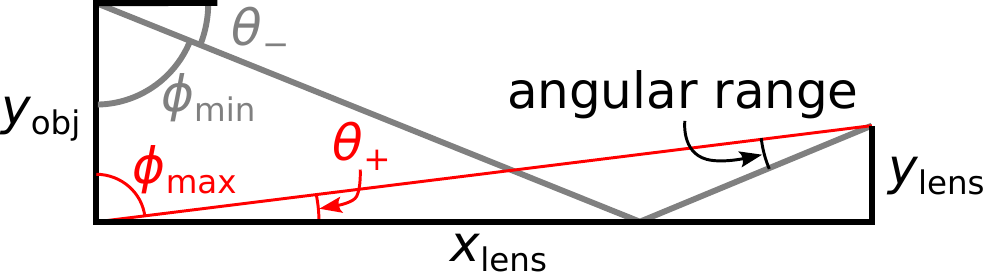}
   \figcaption{The mirage measurement geometry, showing the object at left, lens at right, and the mirage angular limits given by $\theta_{-}$ to $\theta_{+}$.}
   \label{fig:min_max}
\end{center}

The minimum ray angle $\theta_{-}$ can be calculated from the Snell law, from first calculating minimum angle $\phi\subrm{min}$ with respect to the surface normal that still allows inflection,
\begin{equation}
   \phi\subrm{min} = \arcsin(n\subrm{hot} / n\subrm{cool}) \, ,
\end{equation}
and then transferring the reference from the surface normal to the horizontal:
\begin{equation}\label{eq:theta_minus}
   \theta_{-} = (\pi/2) - \phi\subrm{min} \, .
\end{equation}
The minimum ray angle is therefore driven entirely by the temperature difference $\Delta T = T\subrm{hot} - T\subrm{cool}$.

The range $\theta_{-}$ to $\theta_{+}$ therefore gives the angular extent of the mirage, illustrated in \figref{fig:angular_limits} for different values of $\Delta T$ and a wide range of camera distances. For this figure, we have fixed the height of observation at $y\subrm{lens} = \SI{50}{cm}$. The vertical arrows indicate the angular range from $\theta_{-}$ to $\theta_{+}$ for specific choices of camera distance. Thus, we can say that for a lens height of \SI{50}{cm} and $\Delta T = \SI{3}{\celsius}$, the mirage will not be visible for camera distances less than \SI{230}{m}. For a camera distance of \SI{1000}{m}, however, the mirage will be visible at angles between $-0.029\degree$ and $-0.125\degree$. (The 0\degree angle indicates the physical horizon.)

\begin{center}
   \includegraphics[width=0.8\linewidth]{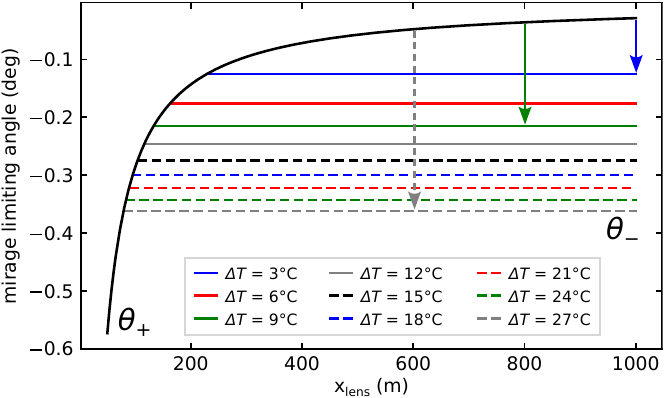}
   \figcaption{The angular limits of the mirage, given as the range from $\theta_{-}$ to $\theta_{+}$ for various values of $\Delta T$ and a range of camera distances. Each of the curves assume a lens height of $y\subrm{lens} = \SI{50}{cm}$. Note that the zero angle here is the horizon angle.}
   \label{fig:angular_limits}
\end{center}

Ruiz reports a measured mirage angle of $\theta_{-} = -0.34\degree$~\cite{Ruiz2019}, which agrees well with the curves shown in \figref{fig:angular_limits} (and which estimate $-0.36\degree$). Although Ruiz used a lens height of $y\subrm{lens} = \SI{100}{cm}$, this only affects $\theta_{+}$ and not $\theta_{-}$.

Vollmer makes the curious assertion that the angular size of mirages is usually about 0.5–-1\degree~\cite{Vollmer2009}, but this surely must be assuming specific (and very favorable unstated) measurement conditions. In fact, \figref{fig:angular_limits} would seem to indicate that $\Delta T > \SI{50}{\celsius}$ is necessary to achieve a mirage angle of 0.5\degree. There does not seem to be a plausible physical situation in which a full 1\degree mirage is possible.

We can also note that if one is willing to place the lens quite close to the ground, then the conditions in which mirages become visible are greatly relaxed. This increases $\theta_{+}$ to nearly zero (the horizon angle), so that situations with small $\Delta T$, such as a roadway on an overcast day, are still sufficient to show a visible mirage, as we show in Sec.~\ref{sec:measurements}

Figure~\ref{fig:min_max} also shows that from $\phi\subrm{min}$ we can also determine the maximum object height that is visible inside the mirage image:
\begin{equation}
   y\subrm{obj,max} = [x\subrm{lens} \tan(\phi\subrm{min})] - y\subrm{lens} \, .
\end{equation}
This naturally scales with the camera distance, but allows a quick estimate of what objects may be visible within the mirage image.


\section{Comparing measured mirage images to the simulation}\label{sec:measurements}

Figure~\ref{fig:meas1} shows a photograph captured at camera heights of $y\subrm{lens} = 110$ and \SI{10}{cm} for a camera-object distance of approximately \SI{480}{m}. At the time of the photographs, the air temperature was approximately $T\subrm{cool} = \SI{30}{\celsius}$, and the sunlight pavement $T\subrm{hot} = \SI{35}{\celsius}$. The camera used to capture the images is a Sony $\alpha$-5100, with a \SI{210}{mm} $f/6.3$ lens~\cite{tolerances,Singer2026}. Since the camera sensor has \SI{3.91}{\um} pixels, an individual pixel's field of view is therefore 0.00107\degree. 

\begin{center}
   \includegraphics[width=0.95\linewidth]{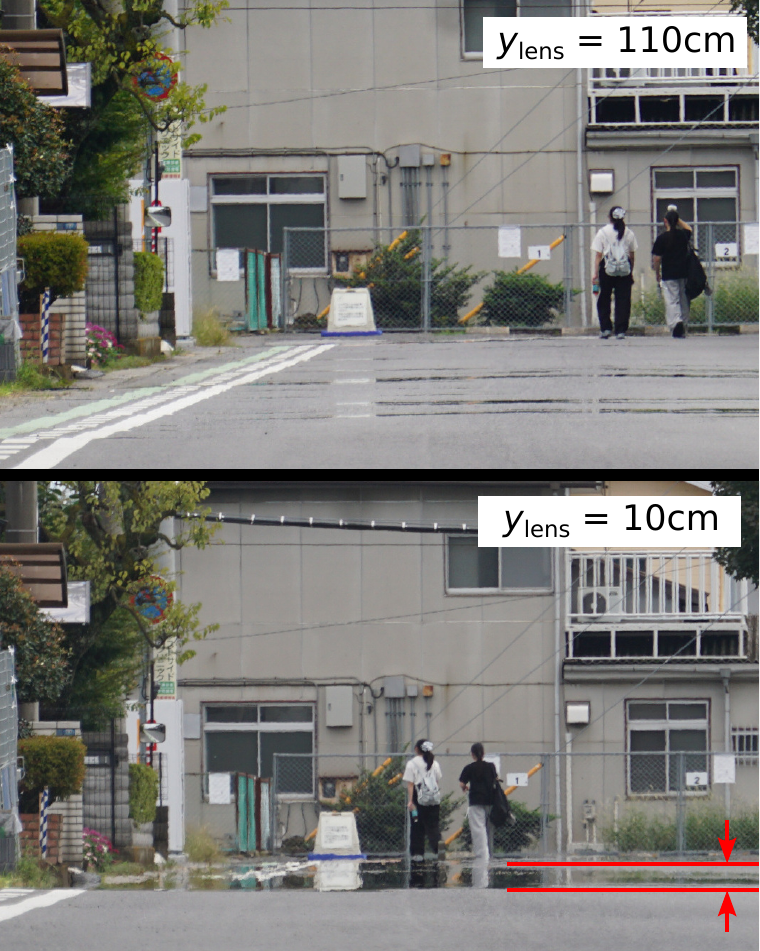}
   \figcaption{Example mirage photographs of the same scene, taken at heights of $y\subrm{lens} = \SI{110}{cm}$ and \SI{10}{cm}, of an urban roadway in Utsunomiya, Japan. (See Visualization 1 for a video that includes vehicles moving through the scene.)}
   \label{fig:meas1}
\end{center}

In the image taken from the high camera height, the mirage is apparent in only thin lines across the pavement. Thus, this camera height is evidently at the upper limit of the mirage visibility for this road under these conditions, though the limiting factor appears to be the flatness of the road rather than the mirage size per se. At the lower camera height ($y\subrm{lens} = \SI{10}{cm}$), the mirage is easily apparent, and we can try to compare measurements with the estimates given by Eqs~\ref{eq:theta_plus}\,\&\,\ref{eq:theta_minus}, which give 
$\theta_{+} = -0.01\degree$ and $\theta_{-} = -0.17\degree$. The red lines superposed on the lower image show the estimated mirage region, which are 48 pixels, and thus 0.051\degree, apart. This is 1/3 of the estimated $\Delta \theta = \big| \theta_{+} - \theta_{-} \big| = 0.15\degree$, but an accurate measurement would also require having a roadway that is flat to much better than this angle --- something which we have found difficult to find here in Japan.

In the supplemental data, we also show a video of a bus, a motorcycle, and a bicycle moving through the mirage region of the image. This is particularly interesting because one can see how the forced airflow from the bus causes a shift in the air temperature above the roadway, and a weakening of the mirage, whereas the motorcycle and bicycle do not.



In any given mirage image, the mirage itself appears like a mirror reflection of the object, and replaces the direct image rather than the two being superposed. From the mirage ray traces, it is not immediately apparent what happens to the portion of the image that is thus replaced. In order to explore this, we have modified the mirage simulation to include light rays emerging from the surface itself. These represent light scattered in all directions by the surface. Figure~\ref{fig:scatter} shows the result, where object rays are shown in black, blue, and green, while rays scattered directly from the surface are shown in a red-orange-yellow color gradient.

\begin{center}
   \includegraphics[width=0.95\linewidth]{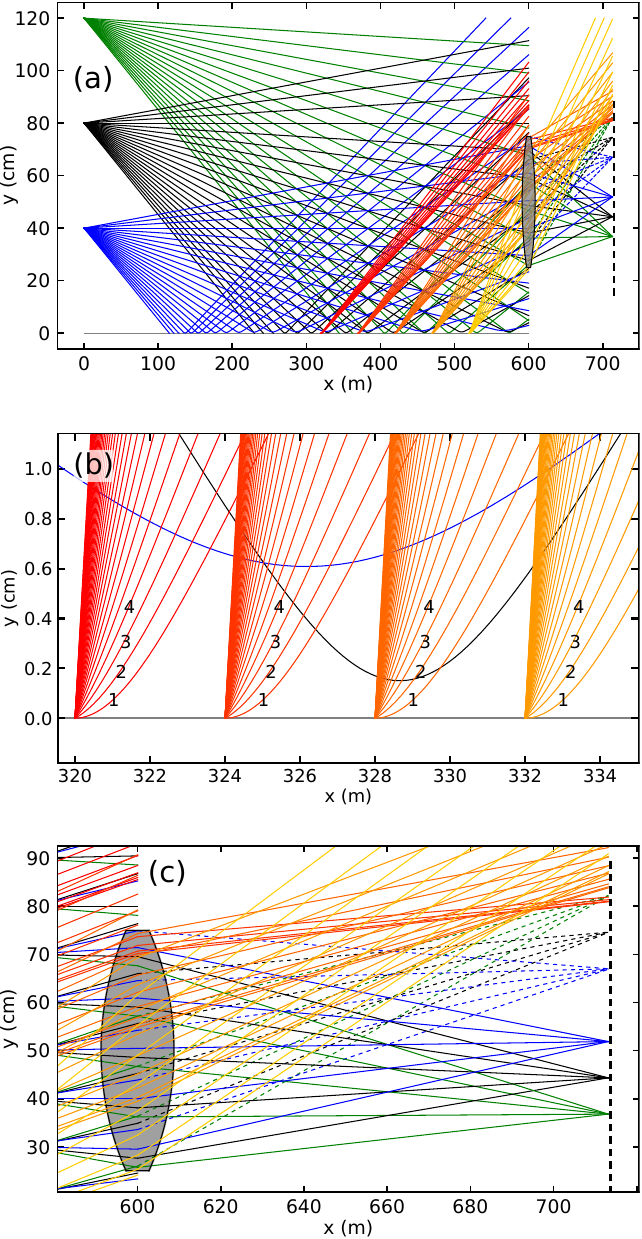}
   \figcaption{A mirage raytrace with $T\subrm{cool} = \SI{43}{\celsius}$, $\beta = \SI{1.0}{cm}$. (a) The red-orange-yellow rays indicate scattered light emerging directly from the surface; (b) a closeup view of the rays emerging from scattering points on the surface; (c) a closeup of the focal region, showing the defocusing of the scattered rays.}
   \label{fig:scatter}
\end{center}

In addition to the mirage rays and direct rays, \figref{fig:scatter}(a) also shows rays emerging from the surface at $x$-locations 32, 37, 42, 47, and \SI{52}{m}. In the closeup view of \figref{fig:scatter}(b), we show rays emerging from the surface at $x$-locations of 32.0, 32.4, 32.8, and \SI{33.2}{m}, and with a wider spread of angles than shown in (a). For each surface point, we model a fan of rays that are uniformly distributed in angle, with the first ray (labelled as ``1'') starting at horizontal, and subsequent rays (\#2, 3, 4, ...) incrementally increasing in angle by 0.01\degree.

As we see from \figref{fig:scatter}(b), even though ray \#1 begins its journey horizontal, it is quickly redirected upward by the strong gradient index at the surface. Subsequent rays, starting at slightly larger initial angles, are deflected to even larger final angles than ray \#1's. Although the bending angle for ray \#1 looks quite large in this view, the physical angle in this case is $\theta\subrm{min} \approx 0.18\degree$, and thus is exaggerated by the usual difference in $x$- and $y$-scales. The deflection of ray \#1 and all subsequent rays creates a void region: even though the rays began their journey in all directions, the index gradient bends them such that the ray bundle emerges from the surface with a minimum angle. For those scatter (red-orange-yellow) rays that are captured by the lens, the deflection has pushed them entirely away from the direct image region (where the rays would reach if undeflected), and even pushes them beyond most of the mirage image region (see the closeup view in \figref{fig:scatter}(c)). These scatter rays are also severely defocused, and so will appear as stray light, in the region at the bottom of a mirage photograph. Moreover, we can see that the lowest angle rays (the ones having the smallest angles $\theta$) within each bundle of scattered rays are more widely separated in angle than are the higher-initial-angle rays. Thus, these low-angle rays will contribute only dim light to the image plane.

Another feature that we can see in \figref{fig:scatter}(c) is that the scattered rays are generally redirected upward in the image plane (which corresponds to the lower region of a mirage photograph). While most of the rays do not overlap with the mirage rays, there is a small overlap area close to the end of the mirage. In the setup modelled here, \SI{120}{cm} is the highest object height which successfully reaches the image plane, $y\subrm{obj,max}$. For objects higher than this, none of the inflected rays will pass through the center of the lens aperture. Thus, we can say that the overlap between the mirage image and the direct image of the pavement will not only be very small, but the scattered rays at the overlap will also be very dim. These properties explain why we see the mirage as a clean reflection image 
rather than an inverted image superposed on top of the direct image.

%

\section{Conclusion}

A detailed raytrace simulation of the mirage allows us to explore some of the lesser-known features of the mirage, and to quantify the angular extent under which the mirage can be visible, Eqs~\ref{eq:theta_plus}\,\&\,\ref{eq:theta_minus}. In addition, we have shown that the mirage image is subject to additional aberration-induced blurring that the direct image is not, and that this blurring is not uniform across the face of the mirage: it is worst where the mirage image meets the direct image. Finally, we show how the scattered rays emerging from the pavement surface itself are deflected by the same gradient that the mirage rays are exposed to, to the extent that the scattered rays are pushed out to the edge of the image. While they overlap the mirage there, the scattered rays will be dim and difficult to detect.

In the supplement to the paper we include a set of Python files for simulating mirage models, and for generating the figures shown in the manuscript. We also include an example mirage video that demonstrates how the mirage changes dynamically with changes in the thermal environment.



\begin{thebibliography}{28}
\bibitem[1]{permissions}
    While we would like to provide examples of such illustrations, we were unable to obtain permission to include them here.
\bibitem[23]{lapserate}
    Note that the thermal decay constant $\beta$, which has units of centimeters, is different from the atmospheric thermal lapse rate, which is intended to model the change in air temperature over kilometers of height.
\bibitem[27]{tolerances}
    Note, however, that the accepted focal length tolerance for commercial lenses is significant, at $\pm 6\%$.
\end{thebibliography}

\end{multicols}

\end{document}